%% file: aanda.tex
\documentclass[letter]{aa}  

\usepackage{graphicx}
\graphicspath{{./Figures/}}
\usepackage{txfonts}
\newcommand{\Gaia}{{\it Gaia }}

\begin{document}

   \title{Scratching the surface with TGAS and RAVE: disk moving groups in the Solar neighborhood}
  \titlerunning{Moving groups in the Solar neighborhood}
   \author{H.H. Koppelman,
        T. Virginiflosia, L. Posti, J. Veljanoski, \and A. Helmi
          }

   \institute{Kapteyn Astronomical Institute, University of Groningen, Landleven 12, 9747 AD Groningen, The Netherlands\\
              \email{koppelman@astro.rug.nl}
}
   \date{}
  \abstract
   {There is long tradition extending more than a century on the identification of moving groups in the Solar neighborhood. However, with the advent of large kinematic surveys, and especially of the upcoming \Gaia data releases there is a need for more sophisticated and automated substructure finders.}
   {We analyze the TGAS$\times$RAVE dataset to identify moving groups in the Galactic disk. These groups of stars may then be used to map dynamical and star formation processes in the vicinity of the Sun.}
   {We use the ROCKSTAR algorithm, a "friends-of-friends"-like substructure finder in 6D phase-space, and analyze the Hertzsprung-Russell diagrams of the groups identified.}
   {We find 125 moving groups within 300 pc of the Sun, containing on average 50 stars, and with 3D velocity dispersions smaller than 10 km/s. Most of these groups were previously unknown. Our photometric analysis allows us to isolate a subsample of 30 statistically significant groups likely composed of stars that were born together. }
   {}

   \keywords{Stars: kinematics and dynamics --
                (Galaxy:) open clusters and associations: general -- (Galaxy:) solar neighborhood
               }

   \maketitle
%

\input{Introduction}

\input{Data}

\input{Results}

\input{Conclusions}

\begin{acknowledgements}
   We thank Joao Alves for many interesting and inspiring discussions. We gratefully acknowledge financial support from a VICI grant from the Netherlands Organisation for Scientific Research, NWO and from the Netherlands Research School for Astronomy, 
NOVA. This work has made use of data from the European Space Agency (ESA)
mission \Gaia (\url{http://www.cosmos.esa.int/gaia}), processed by the 
\Gaia Data Processing and Analysis Consortium (DPAC,
\url{http://www.cosmos.esa.int/web/gaia/dpac/consortium}). Funding for the DPAC
has been provided by national institutions, in particular the institutions
participating in the \emph{Gaia} Multilateral Agreement. 
  from NOVA (the Netherlands School for Astronomy).    
\end{acknowledgements}

\bibliographystyle{aa}
\bibliography{bibliography}

\appendix
\section{Table listing groups and member stars}
\input{table}


\end{document}

%% file: Introduction.tex
\section{Introduction}

The \Gaia dataset \citep[see e.g.][]{brown2016} may constitute the perfect example that proves that an increase in data quality, in terms of number and precision, inescapably leads to unveiling (sub)structure. Here (sub)structure can take many forms such as new (dynamical) components, new (sub)classes of objects, and moving groups. The (upcoming) \Gaia catalogs will yield increases factors of 100 to 1000 in size and in precision. This makes such datasets truly transformational and may result in a revolution in our understanding of the Galaxy, from the scales of star formation, to the dynamics of galaxies, with implications also for cosmology \citep{2001A&A...369..339P}.

\Gaia is unique in its ability to measure the motions of stars in our Galaxy. This information can be used, for example, to identify moving groups. Such moving groups can have multiple origins: dynamical \citep[due to resonances with e.g. the bar or spiral arms,][]{Antoja2012}, associated to star formation events \citep[stars born in the same molecular cloud,][]{dezeeuw1999}, or even to merger events \citep[][and references therein]{2017Galax...5...44J}. The advent of TGAS \citep[Tycho-\Gaia Astrometric Solution,][an appetizer to the upcoming \Gaia second data release, DR2]{Lindegren2016}, already requires the need to use sophisticated and automated methods for finding such substructures \citep{2017A&A...608A..73K, Gagne2018}. 

Here we present the results of applying the substructure finder ROCKSTAR \citep{Behroozi2013} to the catalogue obtained by matching TGAS to the RAVE dataset \citep[RAdial Velocity Experiment,][]{Kunder2016}. ROCKSTAR was originally built to identify (sub)halos in cosmological N-body simulations using a "friends-of-friends"-like algorithm in phase-space. As we will see below, ROCKSTAR is able to identify a plethora of structures in the nearby Galactic disk, whose origin at this point is not fully clear. The far-reaching goal of this work is to understand how star formation has proceeded through the Galactic disk via the identification of kinematic substructures, and to link this information to its dynamical history. This Letter may be seen as a ``proof of concept" of what will be possible with the upcoming \Gaia DR2.

%% file: Data.tex
\section{Data and methods}
The TGAS catalog provides a 5D astrometric solution for $> 2,000,000$ sources by combining Gaia DR1 and the Tycho/Hipparcos catalogs. It has been cross-matched to the last RAVE data release for complementary radial velocities and spectrophotometric distances \citep{McMillanetal2017}, and to the 2MASS stellar catalog for color information. The result is a set of 
200,297 stars with full phase-space information. 

\begin{figure}
  \centering   
  \includegraphics[width=\hsize]{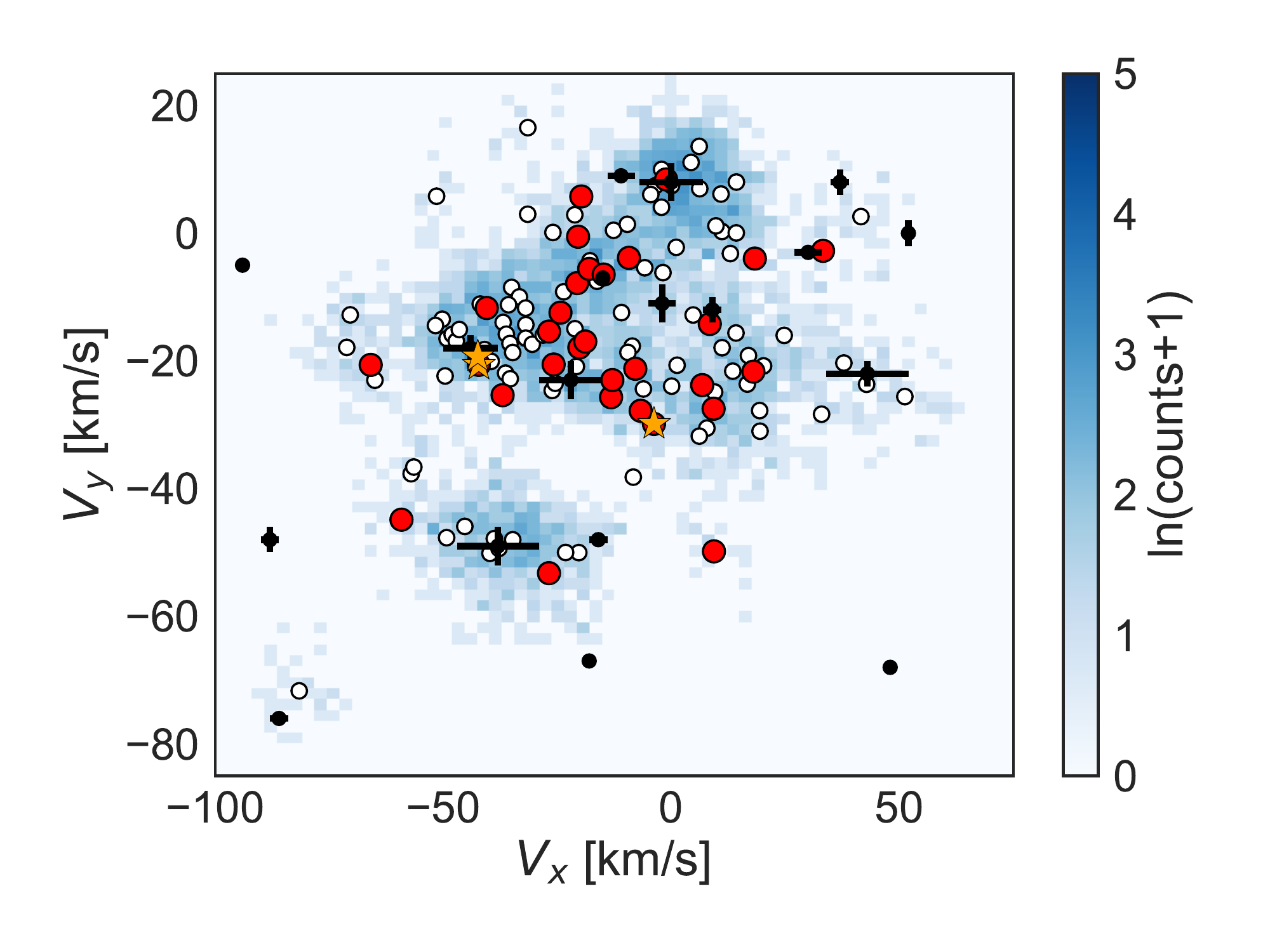}
  \caption{Velocities of stars in our TGAS$\times$RAVE sample. The 125 groups identified in our analysis are marked with circles (colored ones corresponding to the groups in Fig.~\ref{fig:cmds}). The crosses are from \citet{2017A&A...608A..73K}. We have corrected here for the Solar motion with respect to the local standard of rest using 
$(U_\odot,V_\odot,W_\odot) = (11.1 , 12.24 , 7.25)$ km/s \citep{Schonrichetal2010}.}.
     \label{fig:vxvy_dynamic}
\end{figure}

On this dataset we apply several quality cuts, namely we require the error of the line-of-sight velocities to be less than 8 km/s, and that the stars satisfy the following conditions \citep[described in RAVE DR5,][]{Kunder2016} $( (\mathsf{CorrelationCoeff} > 10.0),(\mathsf{SNR\_K} > 20.0),(\mathsf{Algo\_Conv\_K} != 1))$. Our most stringent requirement is on the relative parallax error: $(\epsilon_\varpi / \varpi < 0.1)$. This leaves us with a high-quality sample of 53,328 stars, with more than $90\%$ located within 500 pc from the Sun, and whose  median radial and tangential velocity error is $\epsilon_v \lesssim 1$~km/s. The velocity distribution of these stars is shown in Fig.~\ref{fig:vxvy_dynamic}. 

As stated earlier, we use a modified version of the ROCKSTAR algorithm, which is optimized to identify groups of particles or stars clustered in phase-space. Because ROCKSTAR was conceived for cosmological simulations, we adapt some of its characteristic parameters, and in particular we make it independent of either the mass of the star or the cosmological model\footnote{The most significant modification is the re-scaling of the halo-particle distance metric \citep[see footnote 2 of][]{Behroozi2013}, which we scale by the dispersion in positions (i.e. spatial extent) of the `substructure'. We also remove the dependency on the matter density of the universe $\Omega_m$ in the {\tt AVG\_PARTICLE\_SPACING} parameter.}. Furthermore, in ROCKSTAR's configuration file, we set {\tt FORCE\_RES}=0 (this parameter is related to the simulations' force resolution),
and {\tt BOX\_SIZE}=30~kpc (although this scale is arbitrary as long at it contains all the stars in the set). Furthermore, we fix the minimum size of the structures to be found by ROCKSTAR (defined by {\tt MIN\_HALO\_OUTPUT\_SIZE}, {\tt MIN\_HALO\_PARTICLES}) to 10 stars. Two parameters characterize the clustering algorithm: the {\tt FOF\_LINKING\_LENGTH} $b$, and the {\tt FOF\_FRACTION} $f$. A low value of $b$ will result in structures that are tighter in physical space, while a high value of $f$ leads to more compact structures in velocity space. In this work we explore 3 combinations: $[(b_A,f_A),(b_B,f_B),(b_C,f_C)] = [(0.05,0.4),(0.20,0.8),(0.125,0.6)]$, which we refer to as experiments A, B and C, respectively.

%% file: Results.tex
\section{Results}
\begin{figure*}
  \centering   
  \includegraphics[width=\hsize]{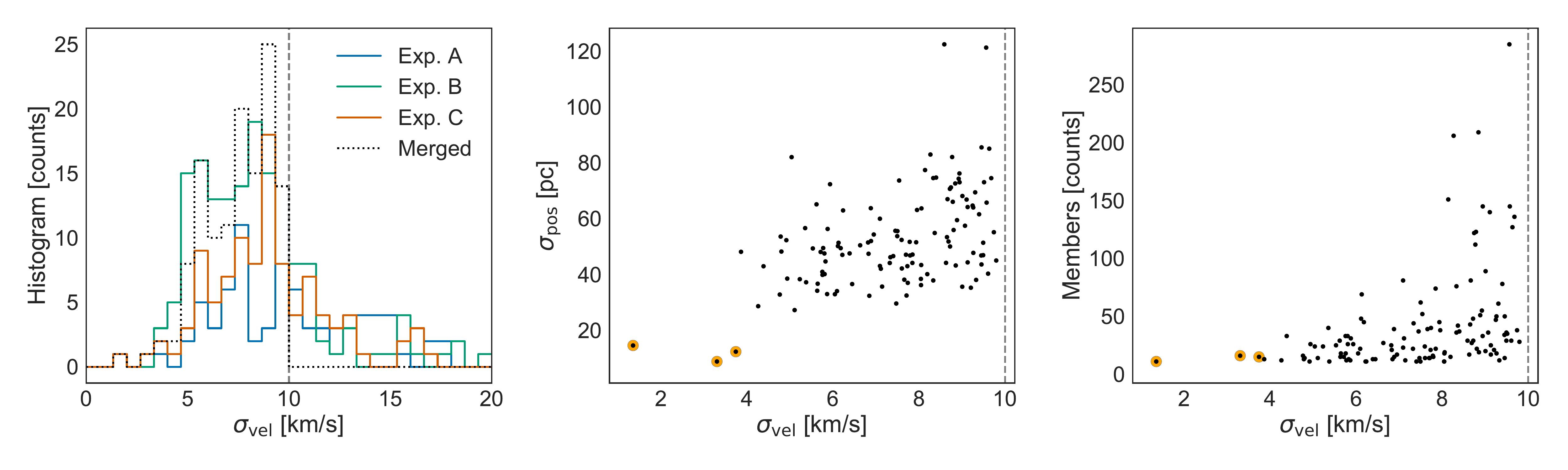}
  \caption{The left panel shows histograms of the velocity dispersion of the groups found in the different experiments. The middle and right panels show the properties of the groups in the merged catalog: number of stars, and dispersions in physical size and in velocity. The latter are computed as 
$\sigma_\mathrm{pos/vel} = \sqrt{\sigma_1^2 + \sigma_2^2 + \sigma_3^3}$, 
where $\sigma_i^2$ are the eigenvalues of the covariance matrix. The median size of the groups is $\sim 50$ pc, and they contain typically a few tens of stars. Groups associated to known star clusters (see Sec.~\ref{sec:previously}) are shown with orange symbols.}
     \label{fig:pos_vel_dispersion}
\end{figure*}

The results of running ROCKSTAR with the above parameters are listed in Table~\ref{table:1}. For each experiment, the algorithm identifies $\sim 100$ substructures, whose distribution of total velocity dispersion is shown in the left panel of Fig.~\ref{fig:pos_vel_dispersion}. We see that this distribution peaks below $\sim 10$~km/s, and has a long tail towards higher values for all experiments. This leads us to make a first cut at 
$\sigma_\mathrm{vel} < 10$~km/s, a value that is well-below the velocity dispersion of the Galactic disk. In what follows we consider only the groups that satisfy this condition. 

We often find significant overlap in the group catalogs produced by the different experiments. Thus, to deal with unique structures we construct a single catalog based on the groups of experiment C and include also groups from experiments A and B that do not overlap at all with those of experiment C. Table \ref{table:1} lists the properties of the merged final catalog, which contains 125 groups hosting a total of $\sim 5000$ stars, i.e. roughly 10\% of the stars in our dataset. 

The central and right panels in Figure \ref{fig:pos_vel_dispersion} show the properties of the merged catalog. We see that most of the groups have $\lesssim 50$ stars, and have an average size of $\sim 50$~pc. Their velocity dispersions range from $5$ to $10$~km/s. 
In both panels, we highlight with orange symbols the groups that can be associated to the Pleiades, Praesepe, and the Hyades (see Sec.~\ref{sec:previously} for details). All three clusters, the Pleiades, Praesepe, and Hyades, are found in all three experiments. Note that they are tighter in space and velocity, although they have notably fewer stars. This could be in part because they were especially targeted by RAVE \citep[see e.g.][]{2014A&A...562A..54C}. Figure \ref{fig:pos_vel_dispersion} shows that the properties of the groups identified by ROCKSTAR appear reasonable. The question that remains is whether they are significant or of a stochastic nature. 

\begin{table}
\caption{Number of structures detected by ROCKSTAR for the three different experiments, together with the number of stars found in those structures. The last row of the table shows the properties of our final (merged) catalog.}             
\label{table:1}      
\centering                    
\begin{tabular}{c c c c}      
\hline\hline                 
exp. \; -- \; ($b,f$)  & groups & groups  & stars \\
 & all & $\sigma_\mathrm{vel} < 10$ km/s & $\sigma_\mathrm{vel} < 10$ km/s  \\
\hline                        
   A \; -- \; (0.05, 0.4) & 157 & 122 & 4805   \\      
   B \; -- \; (0.20, 0.8) & 96  & 42  & 1876   \\
   C  \; -- \; (0.125,0.6) & 103 & 73  & 3992   \\ 
   merged  & 125 & 125 & 5169  \\
\hline                            
\end{tabular}
\end{table}


\subsection{Properties}
 
\begin{figure}
  \centering   
  \includegraphics[width=\hsize]{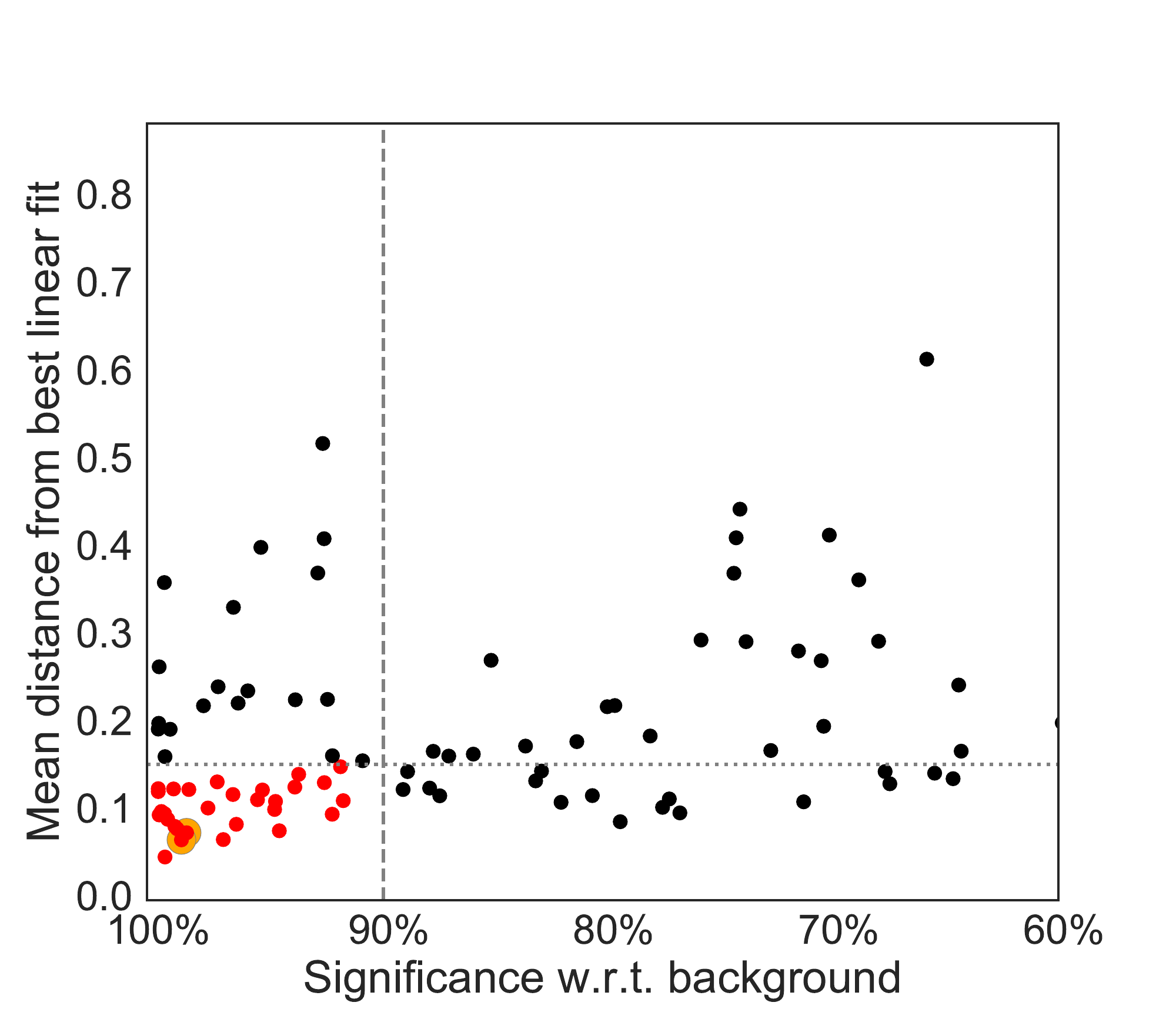}
  \caption{Two statistical properties used to identify the most significant groups: the quality of a straight line fit to the HRD (as measured by the mean distance to the line) against the probability of measuring by chance such a value from a random (background) population of stars. The red dots correspond to groups likely associated to a single star forming site, with the known associations plotted with orange symbols. The dotted line indicates a mean distance of 0.15, while the dashed line is for a 90\% significance level.}
     \label{fig:final_cuts}
\end{figure}
\begin{figure*}
  \centering   
  \includegraphics[width=\hsize]{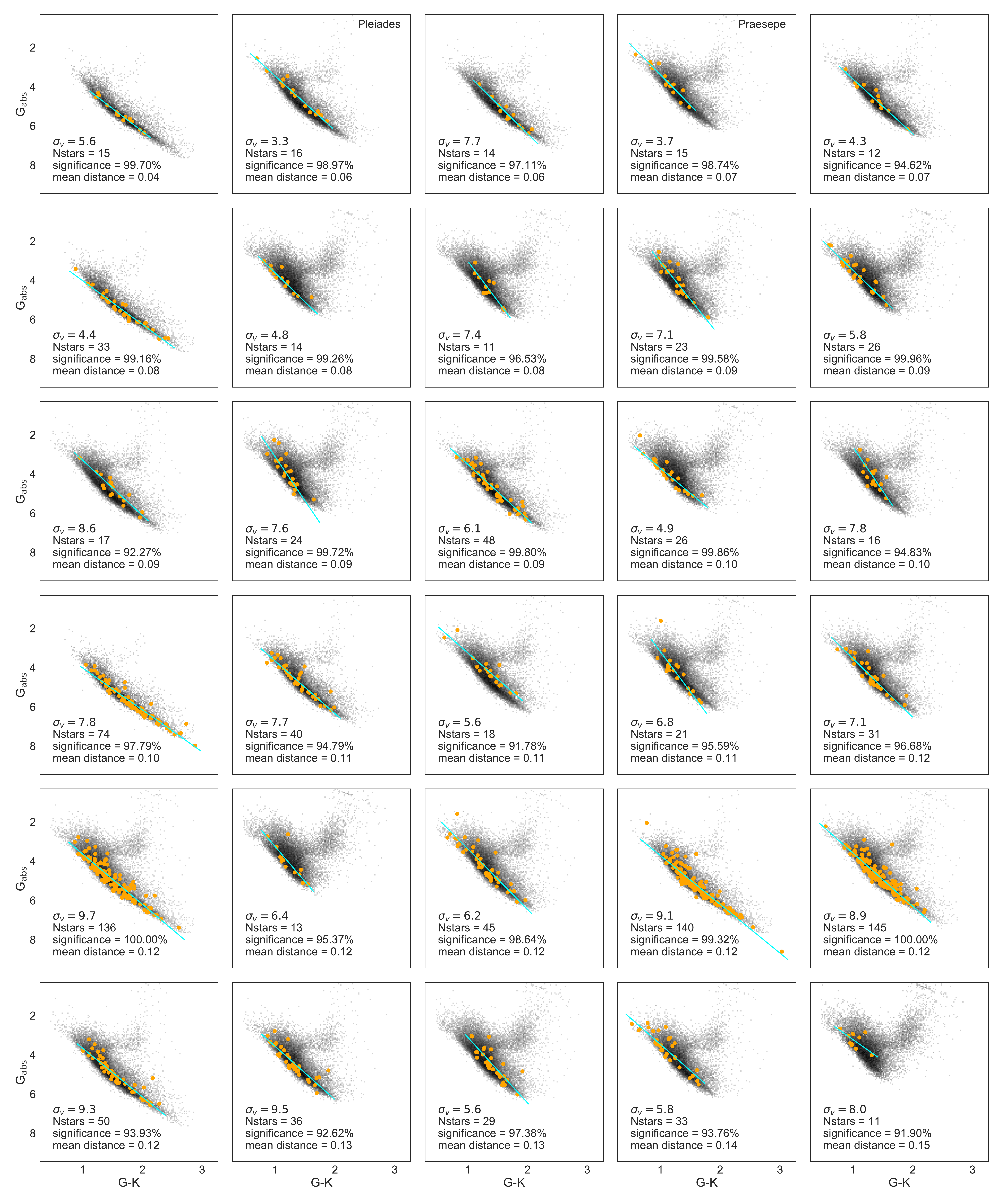}
  \caption{Hertzsprung-Russell diagrams of the 30 selected groups, sorted by their tightness (mean distance) and by their significance with respect to the background. In the left corner of each panel we list the velocity dispersion in km/s, the number of stars in the group, its significance and the mean distance to the straight line fit (which is plotted as a cyan line).}
     \label{fig:cmds}
\end{figure*}

The groups we found could either have a dynamical origin or be composed by stars born together. In the latter case we would expect their Hertzsprung-Russell diagrams (HRD) to be tight and follow the track of a single age and chemical composition stellar population. Since this is relatively straightforward to test, we focus here on assessing which of the groups would be consistent with such an origin using a two-step approach. 

First we fit a straight line to the distribution of stars in the HRD (this is meant to trace the main sequence), and calculate the mean distance of the stars to the line. This is done iteratively: we first we fit a straight line to the entire HRD of the group, next we check with 5 jackknife iterations if we can find a better fit. In every iteration we drop 20\% of the data randomly. The fit resulting in the lowest median distance is stored.

We then calculate the significance of the groups in the HRD by making 10,000 random realizations from a background defined by stars located at a similar distance, with a maximal deviation of 50 pc. Each random HRD is fit like the groups' HRD, again with 5 jackknife iterations. The significance of the group's HRD is then defined to be $1 - N_{\rm ran}/10,000$ where $N_{\rm ran}$ is the number of random HRD that depict a similar or smaller mean distance than that of the group. 

Figure \ref{fig:final_cuts} shows the two statistics plotted against each other. A group needs to be well-fit by a straight line and be significant in comparison to the background to be considered reliable. On this basis, we only consider further groups with a mean distance of < 0.15 and a significance > 90\%. The resulting 30 groups are shown in red in Fig.~\ref{fig:final_cuts}.

The HRD of these 30 groups are shown in Fig.\ref{fig:cmds}, with their stars plotted with orange solid circles (and listed in  Table~\ref{tab:appendix}). For comparison, the different panels include the distribution of background stars with similar distances as the group with a black density map. Overplotted on each HRD is the best fit straight line. Although a straight line is clearly an oversimplification and does not capture all the features seen in the different panels of  Fig.\ref{fig:cmds}, we notice that it nonetheless serves its purpose to identify in a simple manner potentially interesting substructures for further inspection and analysis. Note that we have not plotted here the group associated with the Hyades, and that is because its significance with respect to the background is low (45.3\%), in part because of the low number of member stars we find.

\begin{figure*}
  \centering   
  \includegraphics[width=\hsize]{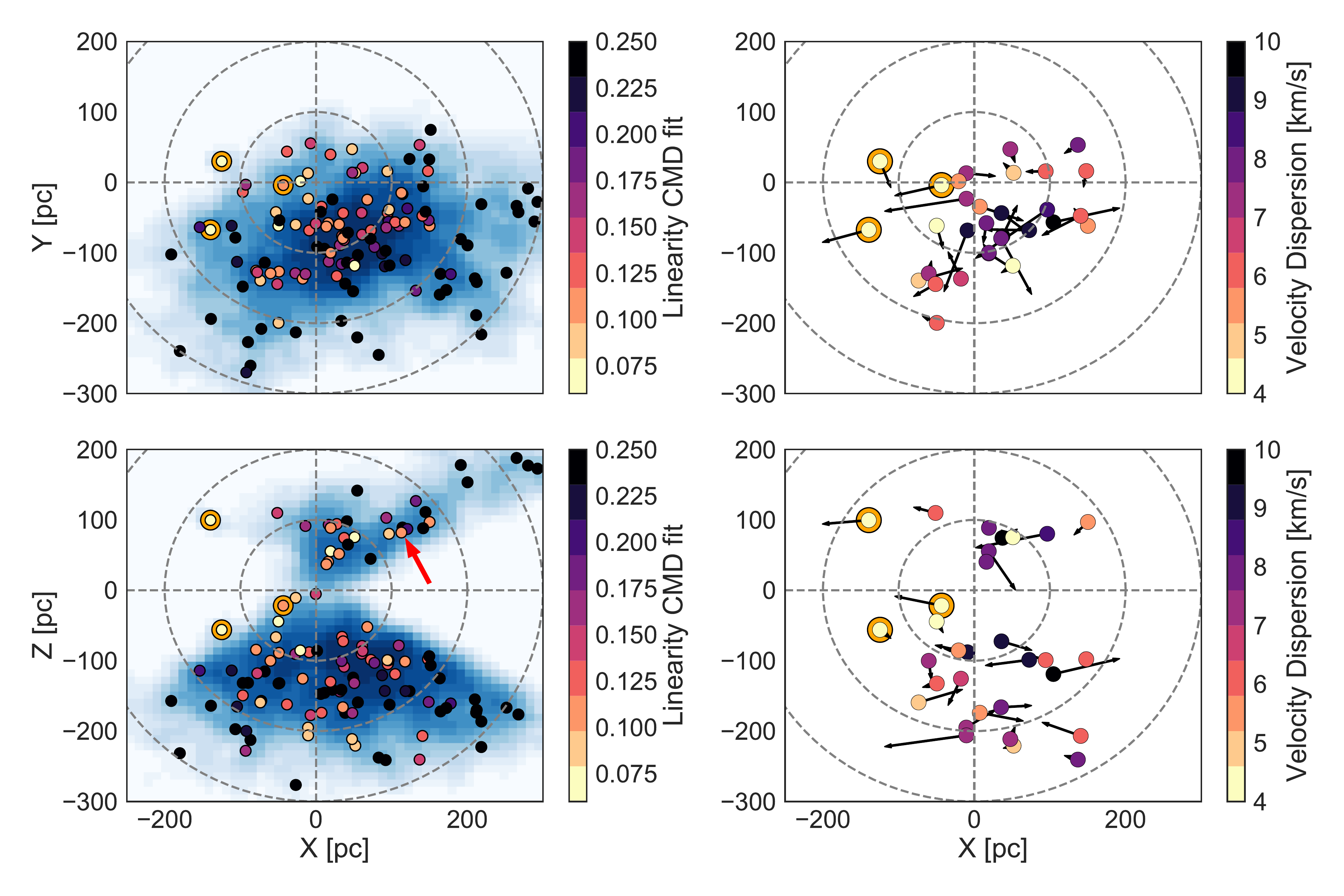}
  \caption{The left panels show the spatial distribution of the 125 groups identified by ROCKSTAR color coded by the statistics used to determine their significance. The panels on the right show the spatial distribution of those 30 that we deem most significant (see Fig.~\ref{fig:cmds}), where the arrows indicate their mean motions, and the colors their velocity dispersion. Detections identified with known moving groups are given by orange symbols. The location of the Scorpius-Canis Majoris stream \citep{Bouy2015} is indicated with a red arrow.}
     \label{fig:2x2XYmaps}
\end{figure*}

The spatial distribution of the groups found is plotted in Figure \ref{fig:2x2XYmaps}. In the two left panels we show the density of the stars in the 125 groups,  whose mean positions are indicated by the colored circles. The lopsided hourglass shape of the density projection on the $XZ$ plane is due to the footprint of RAVE. The structure above the plane, at $Z\sim 80$ pc, overlaps with the stream-like feature composed by OB associations of \citet{Bouy2015}, the Scorpius-Canis Majoris (Sco-CMa) stream. 

The two panels on the right of Figure \ref{fig:2x2XYmaps} show only the mean positions of the 30 photometrically significant groups, where the vectors indicate their mean velocity (corrected for the Solar motion). No clear streaming or coherent motion of the groups is apparent.

\subsection{Comparison with previous work}
\label{sec:previously}
An important question is how many of the groups found are known and how many new discoveries. Besides the tentative association to the Sco-CMa stream, we also explored the catalog of star clusters of \cite{Kharchenko2013}. We found overlap only with the Pleiades and Praesepe, as reported earlier in the paper. The Hyades can also be matched to one of the groups both in 3D positions and in velocities \citep{Riedel2017,Gagne2018}. We also found tentative matches with Blanco 01, Columba, TW Hya, and Upper Scorpius by comparing to the tables in  \cite{Conrad2017,Riedel2017,Gagne2018}. However,  because of the relatively large spatial extent of the structures identified by ROCKSTAR, these matches are preliminary and require a careful statistical analysis of memberships and assessment of contaminants, both of which are beyond the scope of this Letter. Next to these tentative associations, there is some overlap with the kinematic groups found by \cite{2017A&A...608A..73K} and marked with crosses in Fig.\ref{fig:vxvy_dynamic}.

Therefore, overall we have found only a few matches of known structures to our groups. For example, the large cloud of groups below the plane of the disk comprises many unknown structures that cannot be linked to the \citet{Bouy2015} OB associations. We have probably only scratched the surface in terms of the substructure present in the Solar vicinity. The upcoming \Gaia data release will likely reveal a plethora of substructures similar to those found here, whose spatial distribution will be less affected by selection biases (such as the RAVE footprint). It will thus be possible to reveal whether there are any large aggregations of groups, as suggested by \citet{Bouy2015} and hinted at by our analysis as can be seen from the top right panel of Fig.~\ref{fig:2x2XYmaps}.

%% file: Conclusions.tex
\section{Discussion and Conclusions}

By applying the ROCKSTAR substructure finder to the TGAS$\times$RAVE dataset, we have been able to identify 125 moving groups with 3D velocity dispersions smaller than 10 km/s. Through analysis of  their HR diagrams, we have isolated 30 statistically significant groups likely composed by stars born together. The majority of these groups, containing between 10 and 140 stars, were previously unknown. The remaining groups could potentially have a dynamical origin. 

Roughly 10\% of the stars in our sample are in the ROCKSTAR substructures. This fraction is very comparable to the fraction of stars in co-moving pairs found by \citet{Oh2016}, indicating that these co-moving pairs are likely part of the much larger associations found here. 

We find no evidence of the Gould's belt in the spatial distribution of our groups. The distribution is however reminiscent of the large-scale structure reported by \citet{Bouy2015}. Such a ``chain of substructures" possibly indicates how star formation proceeds on the scales of 100 pc and below. On these scales, many  interesting effects come into play: collapse of clouds, feedback from young stars (winds and SN explosions), and interplay with the dynamical field which might lead to compression and shear of the clouds and expansion of the moving groups. A combined study of ages, chemistry and kinematics of the substructures found here, may help us understand how star formation works and how this process is coupled to the Galactic gravitational field.

%% file: table.tex
\begin{table}
\caption{Top 25 members of the 30 groups sorted on mean distance to the best linear fit, as in Fig.\ref{fig:cmds}. The rest of the members of the different groups are available in electronic format.}
\centering
\begin{tabular}{llrrr}
\hline\hline
         &     &           source\_id &      ra &     dec \\
Name & \# &          \Gaia ID   &  $[^{\circ}]$&    $[^{\circ}]$ \\
\hline
group\_115 & 0   & 2334944539380290048 & 359.396 & -25.745 \\
         & 1   & 5076050048450905088 &  41.893 & -25.541 \\
         & 2   & 2343228603581334528 &  10.109 & -27.355 \\
         & 3   & 2416892412308883456 &   4.944 & -14.860 \\
         & 4   & 2457994252899332096 &  21.911 & -12.344 \\
         & 5   & 6528418081683811328 & 351.835 & -46.024 \\
         & 6   & 2457140997516461056 &  22.488 & -13.086 \\
         & 7   & 2493451956706740224 &  35.707 &  -2.822 \\
         & 8   & 2490838726805345280 &  30.569 &  -5.843 \\
         & 9   & 5176541975255450624 &  36.829 &  -8.523 \\
         & 10  & 5139274166070939648 &  26.425 & -19.879 \\
         & 11  & 5171222160043202560 &  40.413 & -11.183 \\
         & 12  & 5171839432742942720 &  39.236 & -11.809 \\
         & 13  & 5026956167075598336 &  17.045 & -32.789 \\
         & 14  & 5171630731692091392 &  36.854 & -12.039 \\
Pleiades & 0   &   65819961494790400 &  58.590 &  24.076 \\
         & 1   &   70941383577307392 &  55.024 &  26.196 \\
         & 2   &   65004707982534016 &  56.724 &  23.583 \\
         & 3   &   66939848447027584 &  57.070 &  25.215 \\
         & 4   &   66471215975411200 &  57.987 &  23.902 \\
         & 5   &   66960258131598720 &  57.574 &  25.379 \\
         & 6   &   66980358578521856 &  57.471 &  25.647 \\
         & 7   &   65150943028579200 &  55.366 &  23.708 \\
         & 8   &   65188085906203520 &  56.272 &  23.702 \\
         & 9   &   68334235349446528 &  55.128 &  24.487 \\
\hline
\end{tabular}

\label{tab:appendix}
\end{table}